\documentclass[%
 reprint,
 amsmath,amssymb,
 aps,
]{revtex4-2}

\usepackage{graphicx}
\usepackage{dcolumn}
\usepackage{bm}
\usepackage{color}
\usepackage{ulem}
\usepackage{amsmath}
\usepackage{wasysym}
\usepackage{epsfig}
\usepackage{braket}
\usepackage[colorlinks=true, letterpaper=true, pdfstartview=FitV, linkcolor=blue, citecolor=blue, urlcolor=blue]{hyperref}
\usepackage{orcidlink}
\usepackage{lipsum}

\begin{document}

\preprint{APS/123-QED}

\title{Non-Hermitian topological superfluidity in a three-dimensional fermi gas with spin-orbit coupling }

\author{Pingcheng Zhu}
\affiliation{Institute for Quantum Science and Technology, Shanghai University, Shanghai 200444, China}
\author{Lihong Zhou}
\email{lihongzh@shu.edu.cn}
\affiliation{Institute for Quantum Science and Technology, Shanghai University, Shanghai 200444, China}
\author{Jianxin Zhong}
\email{jxzhong@shu.edu.cn}
\affiliation{Institute for Quantum Science and Technology, Shanghai University, Shanghai 200444, China}

\date{\today}

\begin{abstract}
The experimental advances in realizing artificial spin-orbit coupling (SOC) and non-Hermitian potentials in ultracold atomic system  open a new avenue for exploring their significant roles in quantum many-body physics. Here, we investigate a non-Hermitian, two-component Fermi system in a cubic lattice with Rashba SOC and complex-valued interaction arising from two-body loss. We adopt the non-Hermitian mean field theory and map out the phase diagram at zero temperature. The interplay of dissipation and on-site interaction drives a dissipation-induced phase transition from superfluid (SF) to normal phase (N). Notably, for weak interaction strengths, this leads to a reentrance of the superfluid state. The presence of SOC significantly expands the parameter regime for both the normal phase and the metastable superfluid phase(MSF). Whereas, the Zeeman field can drive the system from a conventional superfluid into a topological superfluid phase(TSF), characterized by a nontrivial topological invariant. These results enrich our knowledge of pairing superfluidity in Fermi systems.

\end{abstract}

\maketitle


\section{\label{sec:level1}Introduction}

Mostly, the real physical systems are inevitably open quantum systems due to their inescapable coupling to the environment. In recent decades, with the deepening of research on non-Hermitian quantum systems, the exploration of non-Hermitian physics has emerged as a vibrant frontier in modern physics, with profound implications across condensed matter, photonics, and atomic systems. It has been found that such systems can exhibit a wealth of exotic phenomena, such as exceptional points~\cite{ep1,ep2,ep3,pan2}, the non-Hermitian skin effect~\cite{skin1,skin2,skin3,skin4,skin5,skin6,skin7,skin8}, and novel superfluidity phenomena~\cite{soma1,soma2}. Ultracold atomic gases, with their high controllability, provide an ideal platform for studying non-Hermitian physics, where the interaction is tunable through Feshbach resonances~\cite{chin1}, the laser-induced synthetic SOC~\cite{soc1,soc2,soc3,soc4} and the non-Hermiticity can be realized in experiment via laser or electronic beam induced one-body~\cite{skin6,Ott1,Ott2,Ott3,Luo,Ott4,njp}  or two-body dissipation~\cite{Durr1,Takahashi1,Ta1,Takahashi2}. Two-body dissipation refers to a class of decay processes in which two particles are simultaneously lost from the system due to their mutual interaction. Formally, this type of dissipation is described by introducing a complex-valued interaction into the system's Hamiltonian, and the strength of two-body dissipation  can be tuned by  adjusting the depth of the optical lattice~\cite{Durr1,Takahashi1,Ta1,Takahashi2} in experiment. Theoretically, it has been shown that in the few-body regime, the non-Hermiticity arising from two-body dissipation acts to suppress the formation of both two- and three-body bound states~\cite{loss2,zheyu}, which is contrast sharply with other types of non-Hermiticity—such as an imaginary magnetic field or non-Hermitian SOC acting at the single-particle level—which  facilitate the formation of bound states~\cite{zhou1,zhou2,zhou3}. While for the case of many-body physics, the theoretical studies demonstrate that two-body dissipation can induce intriguing forms of fermion superfluidity, including dissipation-induced the phenomenon of superfluid reentrance ~\cite{many1,many2,many3} and phase transitions~\cite{Takahashi1,Takahashi2,phase1,phase2,Tajima2023}, and novel physics~\cite{pan1,pan3}. 

Extensive research has explored the many-body effects of two-body dissipation in both lattice and continuum systems. However, most of these studies focus on the interplay between non-Hermiticity and interactions, the role of SOC and Zeeman fields in such dissipative settings has rarely been addressed. In particular, SOC Fermi gases serve as a unique platform for studying topological superfluidity~\cite{topo0,topo1,topo2,topo3} and exotic pairing mechanisms~\cite{Jiang2011}. Therefore, investigating nontrivial phenomena arising from the interplay among SOC, Zeeman field and non-Hermiticity is highly desirable.

In this work, we consider a  two-component Fermi gas in a cubic lattice with Rahsba SOC, Zeeman field and a complex valued s-wave interaction. Within the framework of a mean-field approach, we determine the ground state of the system at zero temperature and construct the corresponding phase diagram. We demonstrate that such a system exhibits the interesting many-body physics affected by two-body dissipation and SOC. A fundamental feature of this system is the reentrant behavior of the superfluid state under weak interactions. While in lattice systems, either SOC or a Zeeman field alone tends to suppress pairing by increasing the single-particle band gap, their combined effect can be strikingly different. This interplay robustly drives and stabilizes the topological superfluid state. Our results enrich the physical picture and pave the way for understanding the open quantum many-body systems.

This work is organized as follows. In Sec. II we introduce the non-Hermitian effective Hamiltonian of the two-component system and derive the thermodynamic potential at zero temperature limit. We then obtain the gap equation and study the zero-temperature phase diagram in Sec. III.  We first examine the phase diagram in the interaction-dissipation plane in the absence of a Zeeman field. Subsequently, we introduce a Zeeman field and investigate the phase diagram in the dissipation-Zeeman field plane at a fixed interaction strength. Finally, We summarize our work in Sec. IV.

\section{Model Hamiltonian}
We consider a two-component Fermi system with Rashba SOC and a Zeeman field in a three-dimensional(3D) cubic lattice with lattice constant $a$ \cite{Carl2007,Moon2007,Zhai2007}, where particles with opposite pseudospins experience a complex valued s-wave interaction. The Hamiltonian can be written as
\begin{eqnarray}
{ H}=&&\int d\textbf{r}\psi^\dagger(\textbf{r})\left[H_{0}(\textbf{r})-h\sigma_{z}+\lambda(\hat k_{y}\sigma_{x}-\hat k_{x}\sigma_{y})\right]\psi(\textbf{r})\nonumber\\
&&-g\int d\textbf{r}\psi^{\dagger}_{\uparrow}(\textbf{r})\psi^\dagger_{\downarrow}(\textbf{r})\psi_{\downarrow}(\textbf{r})\psi_{\uparrow}(\textbf{r}),
\end{eqnarray}
with 
\begin{eqnarray}
H_{0}(\textbf{r})=-\frac{\hbar^2\nabla^2}{2m}-\mu+V_{L}(\textbf{r})
\end{eqnarray}
here the chemical potential $\mu$ is determined by the total number of atoms $N$ of the system. The optical-dipole  potential $V_L(\textbf{r})=V_0[\cos^2{(k_Lx)}+\cos^2{(k_Ly)}+\cos^2{(k_Lz)}]$ forms the cubic lattice, $V_0$ is the lattice depth and $k_L$ is the recoil momentum giving the lattice spacing $a=\frac{\pi}{k_L}$. The operator $\psi(\textbf{r})=\left[\psi_{\uparrow}(\textbf{r}),\psi_{\downarrow}(\textbf{r})\right]^{T}$ denotes collectively the annihilation operator for spin-up and spin-down atoms.
Here, $h$ is the strength of the Zeeman field and $\lambda$ is the Rashba SOC constant, $\hat{k}_{i}=-i\hbar\partial/\partial i,(i=x,y,z)$ is the momentum operator, $\sigma_{i}$ is the $2\times2$ Pauli matrix, and $g$ represents the 3D interaction constant.

 To capture the low-energy physics in deep lattices, the conventional approach is to expand the field operator in terms of the Wannier functions of the lowest band $\psi_\sigma(\textbf{r})=\sum_{\vec{j}}w_{n=0}(\textbf{r}-\textbf{r}_j)c_{\vec{j}\sigma}$
 (n is the band index, $\sigma$ is spin and $\vec{j}=(j_x,j_y,j_z)$ is 3D lattice site), the mixing between lower and upper bands is neglected. The tight-binding model of the Hamiltonian can be derived straightly
\begin{eqnarray}
{H}=&&-t\sum_{\langle ij\rangle\sigma}c^\dagger_{\vec{i}\sigma}c_{\vec{j}\sigma}-\sum_{\vec{i}\sigma}(\mu+h\sigma_z)c^\dagger_{\vec{i}\sigma} c_{\vec{i}\sigma}-U\sum_{\vec{i}}c^{\dagger}_{\vec{i}\uparrow}c^\dagger_{\vec{i}\downarrow}c_{\vec{i}\downarrow}c_{\vec{i}\uparrow}\nonumber\\
&&+\alpha\sum_{j_x}[(c^\dagger_{j_x\uparrow}c_{j_x+1\downarrow}-c^\dagger_{j_x+1\uparrow}c_{j_x\downarrow})+H.C.]\nonumber\\
&&-i\alpha\sum_{j_y}[(c^\dagger_{j_y\uparrow}c_{j_y+1\downarrow}-c^\dagger_{j_y+1\uparrow}c_{j_y\downarrow})+H.C.]
\end{eqnarray}
where $c_{\vec{i}\sigma}$ ($c^\dagger_{\vec{i}\sigma}$)  is the annihilation (creation) operator of the fermion at site $\vec{i}$. The on-site interaction $U=U_{1}+i\gamma/2$ is  complex-valued, incorporating the interaction strength $U_{1}$ and the two-body loss rate $\gamma$, where $U_1,\gamma>0$. The nearest-neighbor hopping term $t=-\int d^3\textbf{r}w_0^*(\textbf{r}-\textbf{r}_i)[\frac{\hbar^2\nabla^2}{2m}+V_L(\textbf{r})]w_0(\textbf{r}-\textbf{r}_{i+1})$ and the spin-flip hopping terms due to the
Rashba SOC satisfy $\alpha=\lambda\int d^3\textbf{r} w_0^*(\textbf{r}-\textbf{r}_i)\frac{\partial}{\partial x}w_0(\textbf{r}-\textbf{r}_{i+1})$. 

Next, we can express the Hamiltonian of the system in quasi momentum space
\begin{eqnarray}
{H}=&&\sum_{\textbf{k}\sigma}\xi_{\textbf{k}\sigma} c^\dagger_{\textbf{k}\sigma}c_{\textbf{k}\sigma}+\sum_{\textbf{k}\sigma\sigma'}2\alpha(\sigma_x\text{sin}k_y-\sigma_y\text{sin}k_x)c^\dagger_{\textbf{k}\sigma}c_{\textbf{k}\sigma'}\nonumber\\
&&-\frac{U}{N}\sum_{\textbf{k}\textbf{k}'\textbf{q}}c^{\dagger}_{\textbf{k+q/2}\uparrow}c^\dagger_{-\textbf{k+q/2}\downarrow}c_{-\textbf{k}'+\textbf{q/2}\downarrow}c_{\textbf{k}'+\textbf{q/2}\uparrow},
\end{eqnarray}
where $\xi_{\textbf{k}\sigma}=\epsilon_\textbf{k}-\mu_\sigma $,  $\epsilon_\textbf{k}=-2t(\text{cos}k_x+\text{cos}k_y+\text{cos}k_z)$ is the single-particle dispersion in the optical lattice and $\mu_{\updownarrow}=\mu\pm h$ represents the effective chemical potential.

In the following discussion, we take the non-Hermitian mean-feild approximation to the interaction term and consider the case of zero center-of-mass momentum $q=0$. The non-Hermitian mean field order parameters are defined as
\begin{eqnarray}
{\Delta}=&&-\frac{U}{N}\sum_{\textbf{k}}{}_{L}\langle c_{-\textbf{k}\downarrow}c_{\textbf{k}\uparrow}\rangle_{R},\nonumber\\
\tilde\Delta=&&-\frac{U}{N}\sum_{\textbf{k}}{}_{L}\langle c^\dagger_{\textbf{k}\uparrow}c^\dagger_{-\textbf{k}\downarrow}\rangle_{R}.
\end{eqnarray}
where $|E_n\rangle_{R(L)}$ refers to the right (left) eigenstates of Hamiltonian $H$ with energy $E_n(E_n^*)$, and $|E_n\rangle_{R}\neq|E_n\rangle_{L}$, indicating  $\Delta$ and $\tilde{\Delta}$ are not complex conjugate to each other.

Then the mean-feild Hamiltonian in the Nambu basis $\Psi_\textbf{k}=(c_{\textbf{k}\uparrow},c_{\textbf{k}\downarrow},c^\dagger_{\textbf{-k}\uparrow},c^\dagger_{\textbf{-k}\downarrow})^\textbf{T}$ can  be expressed as
\begin{eqnarray}
{H}_{\text{MF}}=&&\frac{1}{2}\sum_{\textbf{k}}\Psi^\dagger_{\textbf{k}}M_\textbf{k}\Psi_\textbf{k}+\frac{1}{2}\sum_{\textbf{k}}(\xi_{\textbf{k}\uparrow}+\xi_{\textbf{k}\downarrow})+\frac{N}{U}\Delta\tilde\Delta
\end{eqnarray}
where the corresponding Bogoliubov Hamiltonian
\begin{widetext}
  \begin{eqnarray}
{M}_{\textbf{k}}=&&\begin{pmatrix}
    \xi_{\textbf{k}\uparrow}&2\alpha(\text{sin}k_y+i\text{sin}k_x)&0 & \Delta\\
  2\alpha(\text{sin}k_y-i\text{sin}k_x)&\xi_{\textbf{k}\downarrow}&-\Delta&0\\
  0&-\tilde\Delta&-\xi_{\textbf{k}\uparrow}&-2\alpha(\text{sin}k_y-i\text{sin}k_x)\\
  \tilde\Delta &0&-2\alpha(\text{sin}k_y+i\text{sin}k_x)&-\xi_{\textbf{k}\downarrow}
\end{pmatrix}
\end{eqnarray}
\end{widetext}
 Once the superfluid is formed, there is a spontaneous $\text{U}(1)$ symmertry breaking and the order parameter obtain a phase. By choosing a specific gauge for which the phase difference of the two order parameters vanishes,  we have  $\Delta=\tilde\Delta=\Delta_{0}$ . The mean-field Hamiltonian can be diagonalized as 
\begin{eqnarray}
{H}_{\text{MF}}=&&\frac{1}{2}\sum_{\textbf{k},\nu=\pm}\left(E_{\textbf{k}\nu}\bar\alpha_{\textbf{k}\nu}\alpha_{\textbf{k}\nu}
+E_{\textbf{k}\nu}\bar\beta_{\textbf{k}\nu}\beta_{\textbf{k}\nu}\right)+\frac{N}{U}\Delta_0^2\nonumber\\
&&+\sum_{\textbf{k}}[\xi_{\textbf{k}}-\frac{1}{2}(E_{\textbf{k}+}+E_{\textbf{k}-})]
\end{eqnarray}
where the quasiparticle energy spectrum is given by $E_{\textbf{k}\pm}=\sqrt{\xi_{\textbf{k}}^2+\Delta_0^2+h^2+4\alpha^2(\text{sin}^2k_x+\text{sin}^2k_y)\pm2E_0}$ with $E_0=\sqrt{\left(h^2+4\alpha^2(\text{sin}^2k_x+\text{sin}^2k_y)\right)\xi_\textbf{k}^2+h^2\Delta_0^2}$. $\bar\alpha_{k\nu}(\bar\beta_{\textbf{k}\nu})$ and $\alpha_{k\nu}(\beta_{\textbf{k}\nu})$ is the quasiparticle (hole) operator for different helicities, which satisfies the anti-commutation relation $\{\bar\alpha_{\textbf{k}\nu},\alpha_{\textbf{k}'\nu'}\}=\delta_{\textbf{k}\textbf{k}'}\delta_{\nu\nu'}$.

To elucidate the many-body physics of the qunantum open system, we  consider the grand partition function
\begin{eqnarray}
\cal{Z}_{\text{MF}}=&&\prod_{\textbf{k},\nu=\pm}(1+e^{-\beta E_{\textbf{k}\nu}})(1+e^{\beta E_{\textbf{k}\nu}})\nonumber\\
&&\times e^{-\beta[\xi_{\textbf{k}}-(E_{\textbf{k}-}+E_{\textbf{k}-})/2]}e^{-\beta N\Delta_0^2/U },
\end{eqnarray}
where $\beta=1/k_{B}T$ denotes the inverse temperature. The thermodynamic potential  can be obtained from $\Omega_{\text{MF}}=-\frac{1}{\beta}\text{ln}\cal{Z}_{\text{MF}}$ 
\begin{eqnarray}
\Omega_{\text{MF}}=&&-\frac{1}{\beta}\sum_{\textbf{k},\nu=\pm}\text{ln}(1+e^{-\beta E_{\textbf{k}\nu}})(1+e^{\beta E_{\textbf{k}\nu}})\nonumber\\
&&+ \sum_{\textbf{k}} \left[\xi_{\textbf{k}}-\frac{1}{2}(E_{\textbf{k}+}+E_{\textbf{k}-})\right]+\frac{N}{U}\Delta_0^2,
\end{eqnarray}

In this work, we consider the limit $\beta\to\infty$ to elucidate the ground-state physics. At zero temperature, the thermodynamic potential reduces to 
\begin{eqnarray}
\Omega_{\text{MF}}=&&\sum_{\textbf{k}}[E_{\textbf{k}+}\Theta(-\text{Re}(E_{\textbf{k}+}))+E_{\textbf{k}-}\Theta(-\text{Re}(E_{\textbf{k}-}))]\nonumber\\
&&+ \sum_{\textbf{k}} \left[\xi_{\textbf{k}}-\frac{1}{2}(E_{\textbf{k}+}+E_{\textbf{k}-})\right]+\frac{N}{U}\Delta_0^2.
\end{eqnarray}

\section{Phase Diagram }
To construct the zero-temperature phase diagram, we derive the gap equation from the stationary condition of the thermodynamic potential $\partial \Omega_{\text{MF}}/\partial\Delta_0=0$, 
\begin{eqnarray}
\frac{N}{U}=\sum_{\textbf{k}}\left(\frac{1+h^2/E_0}{4E_{\textbf{k}+}}+\frac{1-h^2/E_0}{4E_{\textbf{k}-}}\right).
\end{eqnarray}
The order parameters are obtained by solving the gap equation for different choices of interaction strength $U_1$ , dissipation $\gamma$, SOC strength $\alpha$ and Zeeman field $h$.  To distinguish different phases, we resort to a method based on the energy difference between the superfluid state and the normal state as $\Delta E=\Omega_{\text{MF}}(\Delta_0)-\Omega_{\text{MF}}(0)$~\cite{many1,many3}. In the following numerical calculations, we adopt the hopping integral $t$ and the lattice constant $a$ as units of energy and length, respectively. 
\begin{figure}[t]
\includegraphics[width=0.48 \textwidth]{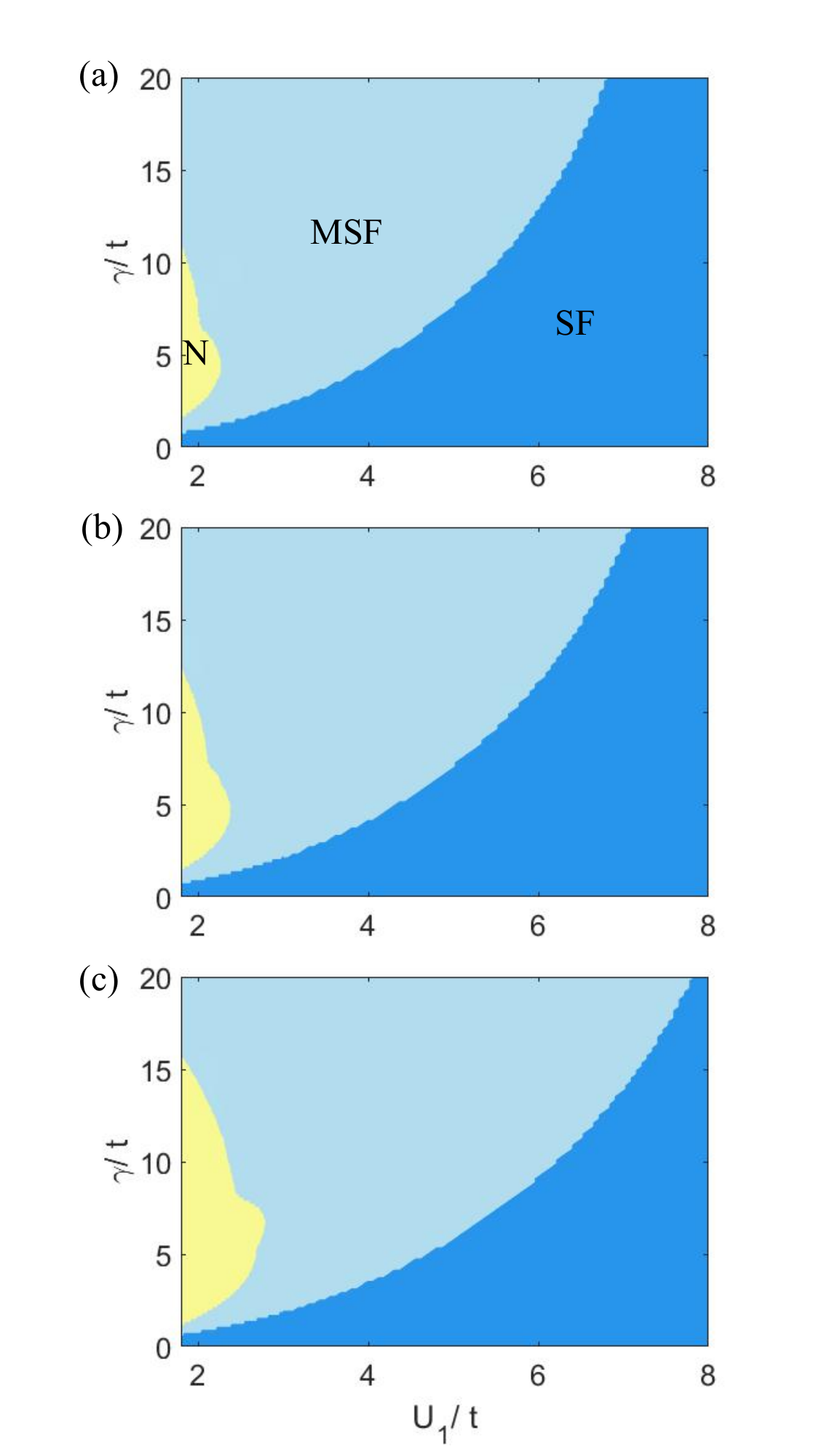}
\caption{Zero-temperature phase diagram in the parameter space of the interaction strength $U_1$ and  the dissipation $\gamma$ with the spin-orbit coupling strength: (a) $\alpha/t=0$, (b)  $\alpha/t=0.4$, and (c) $\alpha/t=0.8$. The blue, light blue, and yellow regions denote stable superfluid, metastable superfluid, and normal states, respectively. In all figures we take $\mu/t=0$, $h/t=0$. Due to numerical constraints, the region with small $U_1$ is not shown.}
\label{fig1}
\end{figure}

 We first investigate the effect of Rashba SOC. In Fig.~\ref{fig1}, we present the mean-field zero-temperature phase diagram in the interaction strength–dissipation plane  $(U_1,\gamma)$ for different SOC strength, and elucidate the intricate interplay between the interaction strength, dissipation, and SOC. The yellow region corresponds to the stable normal phase (N), characterized by $\text{Re}(\Delta_0)=0$ . The blue region represents the stable superfluid phase (SF), distinguished by $\text{Re}(\Delta_0)\neq0$ and $\Delta E<0$, which is an effective ground state of the non-Hermitian Hamiltonian. The light blue region denotes the metastable superfluid phase (MSF), identified by Re$(\Delta_0)\neq0$ and $\Delta E>0$. An interesting phase occurs in the weakly interacting regime. As dissipation increases, the system undergoes a sequence of phase transitions: from the SF phase to the MSF phase, and finally to the normal phase. Upon further increasing the dissipation strength, the system re-enters the MSF phase. This reentrant behavior is a direct consequence of the quantum Zeno effect~\cite{Durr1,Takahashi1}: strong two-body dissipation suppresses intersite particle hopping, favors on-site pair formation, and thereby enhances the superfluid order parameter. The physical mechanism is the same as that discussed in Ref.\cite{many1,many3}. Another characteristic feature  across all three figures illustrated in panels Fig.~\ref{fig1}(a) to (c) is that, as the SOC strength increases — the areas of both the normal and metastable superfluid phases expand, while the region of the stable superfluid phase shrinks. This suppression of superfluidity by SOC can be understood from the single-particle energy spectrum.

\begin{figure}[t]
 \centering
\includegraphics[width=0.48 \textwidth]{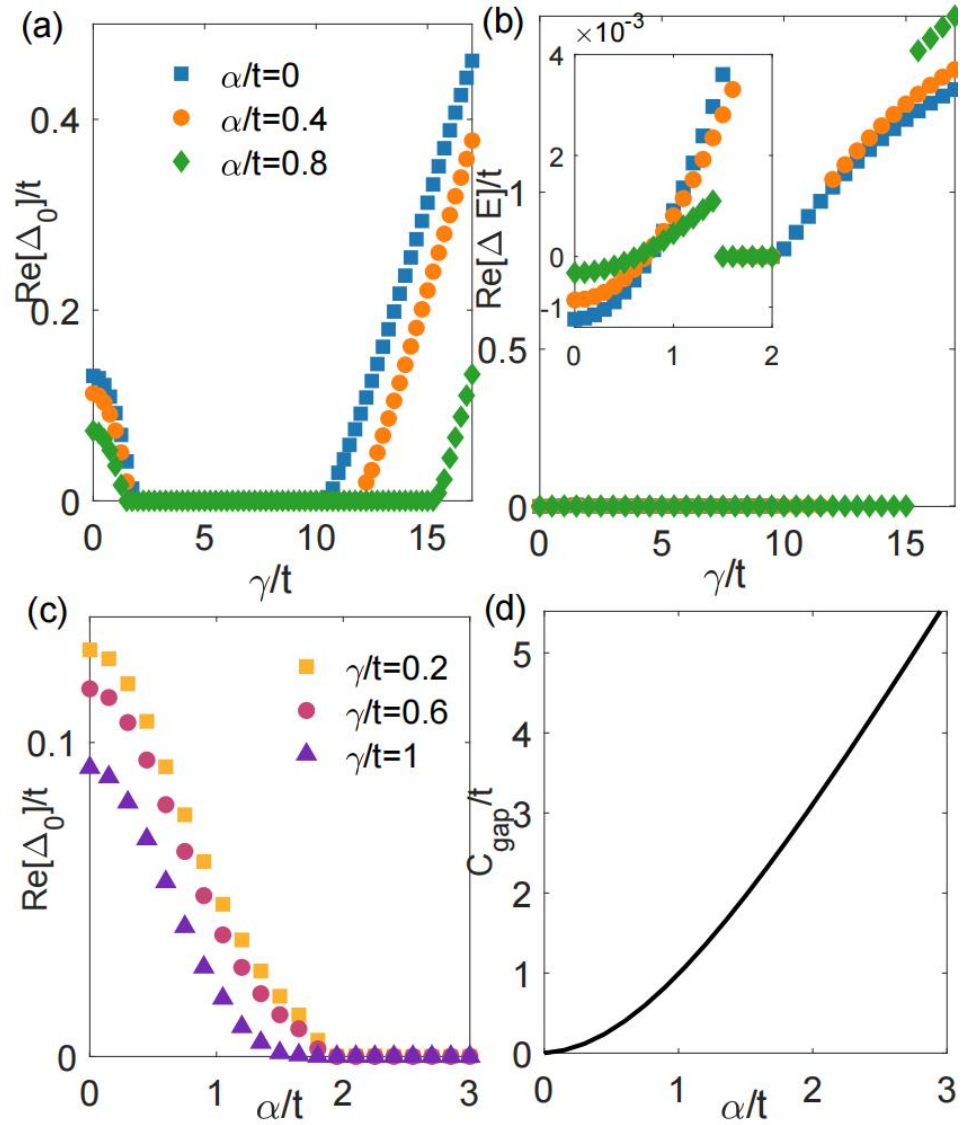}
\caption{ The real parts of (a) order parameter $\Delta_0/t$ and (b)  condensate energy $\Delta E/t$ as functions of the dissipation $\gamma$ for SOC strength $\alpha/t=(0,0.4,0.8)$. The inset in (b) provides a magnified view at low dissipation strengths. (c) The real parts of  order parameter $\Delta_0/t$ as a function of SOC strength $\alpha$ with $\gamma/t=(0.2,0.6,1)$. (d) The energy gap of the single-particle spectrum as a function of the spin-orbit coupling strength $\alpha$. In all figures we take $\mu/t=0$, $U_1/t=1.8$ and $h/t=0$.}
\label{fig2}
\end{figure}

To gain deeper insights into the phase transition in the weak-interaction regime, in Figs.~\ref{fig2} (a) and (b) we present the real part of the order parameter $\Delta_0$ and the condensate energy $\Delta E$ as functions of dissipation $\gamma$ with different SOC strength. As shown in Fig.~\ref{fig2} (a), the real part of the order parameter vanishes as dissipation increases, indicating that the superfluid state is destroyed and the system enters the normal state. However, with further increased dissipation, the superfluid state recovers and the order parameter rises. The corresponding condensate energy starts from a negative value, increases with dissipation, and jumps discontinuously to zero; in the strong-dissipation regime, it turns positive again, indicating the existence of a MSF state, as shown in Fig.~\ref{fig2}(b). In both Figs.~\ref{fig2}(a) and (b), SOC expands the parameter regimes in which $\Delta_0$ and $\Delta E$ vanishes, thereby enlarging the area of the normal state and suppressing the superfluid phase. This inhibitory role of SOC is further illustrated in Fig.~\ref{fig2}(c) and (d), where the order parameter decreases and the energy gap of a single-particle spectrum increases with increasing SOC strength. In the presence of SOC, the single-particle spectrum splits into two helicity branches. The monotonic increase of the energy gap in these two branches implies a reduced probability of inter-branch pairing.

\begin{figure}[t]
 \centering
\includegraphics[width=0.42\textwidth]{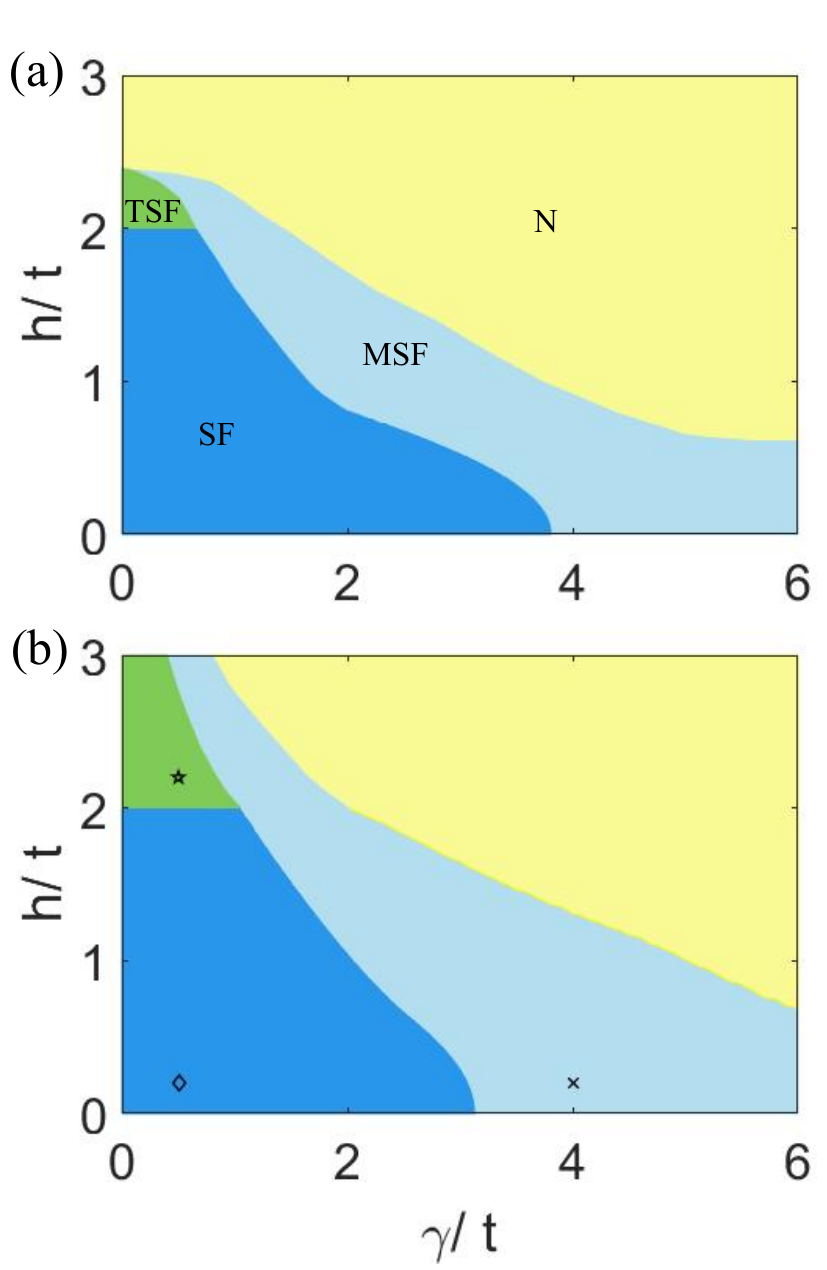}
\caption{Zero-temperature phase diagram in the plane of the dissipation $\gamma$ and Zeeman field $h$ with the SOC strength: {(a) $\alpha/t=0.6$, (b)  $\alpha/t=1$}. The green region shows the topological superfluid phase. In all figures we take $U_1/t=4$ and $\mu/t=0$.}
\label{fig3}
\end{figure}

The Zeeman field is also a commonly used tuning parameter that breaks spin degeneracy, modifies the energy-level structure, and thereby influences the pairing mechanism, phase diagram, and topological properties of the system.  In experiments, the topological band structure has been observed by combinations of optical lattice, Zeeman field and SOC ~\cite{exp1,exp2}. To  elucidate the intricate interplay of dissipation, Zeeman field and SOC, we present the zero-temperature phase diagram in the plane of Zeeman field strength and dissipation $(h,\gamma)$ for two different SOC strength in Fig.~\ref{fig3}. In particular, we identify a new phase: a topological superfluid(TSF) state characterized by a non-zero Chern number (C) in the $k_z$-plane. This integer topological invariant serves as an effective predictor for the topological transition. In Fig.~\ref{fig3}, regions with $C \neq 0$ (indicated in green region) mark the TSF phase. The interplay of SOC and Zeeman field splits the single-particle energy spectrum into distinct spin-mixed branches, enabling effective p-wave pairing and facilitating the emergence of the TSF phase. For weak SOC as show in Fig.~\ref{fig3}(a), the TSF phase is confined to a narrow window between the normal and trivial superfluid phases. Upon increasing SOC, this topological region broadens markedly, accompanied by a concomitant shrinkage of the SF phase. This behavior can be attributed to the Rashba SOC operating through different mechanisms under varying Zeeman field strengths. For a weak Zeeman field, the time-reversal symmetry breaking is not particularly pronounced and the Fermi surface mismatch between spin species is minimal. Here, the dominant pairing channel remains the zero-momentum, spin-singlet-like BCS state which is suppressed as the increase of SOC. For a large Zeeman field, the Fermi surfaces are severely shifted, causing a pronounced Fermi surface mismatch. This strongly suppresses and can completely destroy the conventional BCS pairing, the interplay of SOC and Zeeman field  pushing the system toward a novel phase.  In this regime, the SOC plays a constructive and essential role. Increasing the SOC strength in this regime enhances the efficiency of this alternative pairing channel. It increases the phase space available for forming these resilient pairs, which are immune to the large Zeeman field. This can be viewed as SOC engineering a path to superfluidity, such as a helical superfluid or a topological phase, that explicitly relies on strong spin-orbit interaction to exist. The crossover between these two regimes marks a transition in the dominant pairing mechanism. At a critical Zeeman field, the system moves from a topologically trivial BCS-like superfluid toward a non-trivial superfluid phase. In the latter, the order parameter may even originate from inter-band pairing and exhibit topological properties. Further investigation reveals that at a fixed Zeeman field, there exists an optimal SOC that maximizes its enhancement effect on the pairing order parameter. SOC serves as the cornerstone for designing topological phases, while the Zeeman field acts as a switch for controlling them. Their interplay effect is key to exploring novel topological quantum states and realizing future low-energy electronic devices.
 
\begin{figure}[t]
 \centering
\includegraphics[width=0.47\textwidth]{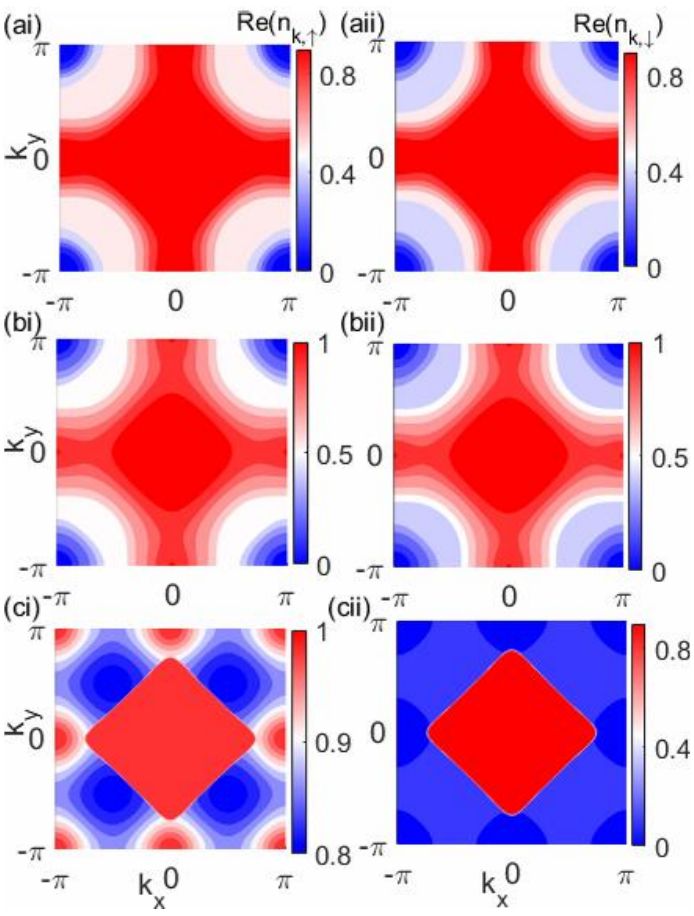}
\caption{The momentum space density distribution in $k_z=0$ plane for three phase marked in Fig. 3(b). The order parameters are (a) $\Delta_0/t=0.855+0.127i$ (superfluid phase), (b) $\Delta_0/t=0.755+1.087i$ (metastable superfluid phase),  (c) $\Delta_0/t=0.016+0.008i$ (topological superfluid phase), other parameters are the same as Fig. 3(b). }
\label{fig4}
\end{figure}

It is well known that topological phases are typically characterized by topologically protected surface or edge states. For lattice models, the phase boundary can be determined by the closure of the excitation gap~\cite{Y.Li-Jun2015}. According to quasiparticle energy spectrum, the excitation energy gap closes at $h=\text{Re}\sqrt{\xi_{\textbf{k}}^2+\Delta_0^2}$ with $k_x=(0,\pm\pi),k_y=(0,\pm\pi)$, corresponding to the following three conditions
 \begin{eqnarray}
 &h_{c}=\text{Re}\begin{cases}\sqrt{(-4t+2t\cos{(k_z )}+\mu)^2+\Delta_0^2}\\
\sqrt{(2t\cos{(k_z )}+\mu)^2+\Delta_0^2}\\
\sqrt{(4t+2t\cos{(k_z )}+\mu)^2+\Delta_0^2}
  \end{cases}
\label{cond}
\end{eqnarray}
Based on the Eq.(\ref{cond}), we can  get the critical Zeeman
 fields at a fixed $k_z$ value.

Complementary insight into the different phases can be gained from the zero-temperature momentum-space density distribution for each spin species, which is obtained from the derivative of the thermodynamic potential with respect to the corresponding chemical potential, $n_\sigma=-\partial \Omega_{\text{MF}}/\partial\mu_\sigma$. In Fig.~\ref{fig4}, we present the momentum distributions of different spin components for SF phase, MSF phase, and TSF phase, respectively. We found that for both SF phase and MSF phase, the momentum distributions of the two spin components exhibit similar trends: higher occupation near zero momentum, decreasing toward the Brillouin zone edge, and displaying 90-degree discrete rotational symmetry due to the cubic lattice, as shown in Fig.~\ref{fig4}(a) and (b). The MSF state more tends to take a central position. In contrast, for TSF phase shown in Fig.~\ref{fig4}(c), the spin-up particles show a dip-like structure in the four symmetric regions inside the Brillouin zone, while the spin-down distribution shows zero occupation near the excitation energy gap close points: as  $(k_x,k_y)=(0,\pm \pi)$,$(\pm\pi,0)$ and $(\pm\pi,\pm\pi)$. This distinctive dip-like feature, unique to the topological superfluid phase, provides a key experimental signature for identifying this state.

\section{Summary}
 In this work, we have systematically studied the many-body phase diagram of a two-component Rashba SOC Fermi gas with two-body loss in a cubic lattice. Our results reveal that with increasing dissipation, the system in the weak interaction regime sequentially enters three distinct phases: superfluid, normal, and metastable superfluid. The intriguing property of many-body phase is the reentrant behavior of superfluid state due to quantum Zeno effect. In addition, the introduction of  Rashba SOC can enlarge the single-particle energy gap, thereby extending the normal phase region and suppressing the superfluid phase  in the phase diagram. More importantly, the coexistence of Rashba SOC and a Zeeman field promotes the emergence of topological superfluid state which is robust against local impurities and disturbances as long as the perturbation is not strong enough to close the bulk energy gap and induce a topological phase transition. It would also be interesting to study the possibility of other exotic pairing phases, such as the finite center-of-mass momentum pairing state and the breached paired phase.
 
\begin{acknowledgments}
This work is supported by  the National Natural Science Foundation of China (Grant No. 12374046), the Shanghai Science and Technology Innovation Action Plan (Grant No. 24LZ1400800), the State Key Laboratory of Micro-nano Engineering Science (Grant No. MES202605).
\end{acknowledgments}

\nocite{*}

\bibliography{apssamp}

\end{document}